\documentclass[showpacs,10pt,twocolumn,prl]{revtex4-1}

\usepackage{amsmath}
\usepackage{amssymb}
\usepackage{graphicx}
\usepackage{amssymb}
\usepackage{graphics}
\usepackage{epsfig}
\usepackage{CJK}
\usepackage{color}
\usepackage{soul}

\begin{document}

\begin{CJK*}{GBK}{}
\title{Thermal transport properties of IrSbSe}
\author{Yu Liu,$^{1,*}$ Milinda Abeykoon,$^{2}$ Niraj Aryal,$^{1}$ David Graf,$^{3}$ Zhixiang Hu,$^{1,4}$ Weiguo Yin,$^{1}$ and C. Petrovic$^{1,4}$}
\affiliation{$^{1}$Condensed Matter Physics and Materials Science Department, Brookhaven National Laboratory, Upton, New York 11973, USA.\\
$^{2}$National Synchrotron Light Source II, Brookhaven National Laboratory, Upton, New York 11973, USA.\\
$^{3}$National High Magnetic Field Laboratory, Florida State University, Tallahassee, Florida 32306-4005, USA.\\
$^{4}$Materials Science and Chemical Engineering Department, Stony Brook University, Stony Brook, New York 11790, USA.}
\date{\today}

\begin{abstract}
We report a thermal transport study of IrSbSe, which crystallizes in a noncentrosymmetric cubic structure with the $P2_13$ space group and shows a narrow-gap semiconducting behavior. The large discrepancy between the activation energy for conductivity [$E_\rho$ = 128(2) meV] and for thermopower [$E_S$ = 17.7(9) meV] from 200 to 300 K indicates the polaronic transport mechanism. Electrical resistivity varies as $exp(T_0/T)^{1/4}$ and thermopower varies as $T^{1/2}$ at low temperatures, indicating that it evolves into the Mott's variable-range hopping dominant conduction with decreasing temperature. IrSbSe shows relatively low value of thermal conductivity ($\sim$ 1.65 W/K$\cdot$m) and thermopower of about 0.24 mV/K around 100 K, yet poor electrical conductivity. On the other hand, high vacancy defect concentration on both Ir and Sb atomic sites of up to 15\%, suggests high defect tolerance and points to possibility of future improvement of carrier density by chemical substitution or defect optimization.
\end{abstract}

\maketitle
\end{CJK*}

\section{INTRODUCTION}

New transition metal chalcogenides may enable transformative changes in thermoelectric energy creation and conversion \cite{RoychowdhuryS,OuyangY,BanikA}. Ternary CoSbS with Kondo-insulator-like intrinsic magnetic susceptibility features high value of thermopower up to 2.5 mV/K at 40 K but also relatively large value of thermal conductivity $\sim$ 100 W/K$\cdot$m near the thermopower peak temperature \cite{DuQ}. This inhibits the thermoelectric figure of merit $ZT= (S^{2}$/$\rho$$\kappa$)$T$, where $S$ is thermopower, $\rho$ is electrical resistivity, $\kappa$ is thermal conductivity, and $T$ is temperature.

Electronic correlations are important in many transition metal and rare earth-based materials \cite{PallsonG,KoshibaeW}. Theoretical study confirms that the colossal thermopower in CoSbS is due to carrier correlation and large increase of effective mass \cite{GuptaR}. In addition, it was suggested that thermoelectric power factor in CoSbS may show 10$^7$ times increase in magnetic field due to high spin-orbit coupling (SOC) effect on thermopower \cite{GuptaR}. The SOC causes considerable enhancement of electrical conductivity, which is rather sensitive to Co atomic moment and yet leaves the phonon dispersion and thermal conductivity unaffected.

Iridium-based materials with strong SOC host a variety of exotic quantum phases but also properties of interest for applications \cite{RauJ,TakagiH,WanX,YL}. In IrBiSe, for example bulk electronic bands are split by giant spin orbit splitting about 0.3 eV and are fully spin polarized \cite{LiuZ}. Electronic states in IrBiSe with three-dimensional (3D) chiral spin texture with negative and positive chiralities along crystallographic [111] direction are of interest for spin sensor applications and could exhibit spin-triplet superconductivity upon doping \cite{LiuZ}.

In this work we report the thermal transport properties of noncentrosymmetric and cubic IrSbSe, isostructural to IrBiSe. We observed vacancy defects on both Ir and Sb atomic sites, i.e. Ir$_{0.90}$Sb$_{0.85}$Se stoichiometry, leading to low values of thermal conductivity. The large discrepancy between the activation energy for electrical conductivity and for thermopower from 200 to 300 K suggests the polaronic transport mechanism. With decreasing temperature, it evolves into the Mott's variable-range hopping dominant mechanism at low temperatures. Whereas high values of electrical resistivity inhibit $ZT$, relatively high tolerance of the crystal structure to defect formation may allow for further carrier tuning and optimization of thermoelectric performance, similar to IrBiSe \cite{LiuZ}.

\section{EXPERIMENTAL DETAILS}

IrSbSe polycrystal was synthesized via solid state reaction starting from an intimate mixture of high purity elements Ir powder (4N, Alfa Aesar), Sb and Se pieces (5N, Alfa Aesar) with a stoichiometric ratio. The starting materials were mixed and ground in an agate mortar, pressed into a pellet and sealed in an evacuated quartz tube backfilled with pure argon gas. The tube was heated to 500 $^\circ$C and dwelled for 12 h, and then slowly heated to 800 $^\circ$C and reacted for 5 days followed by furnace cooling. The chemical composition was determined by multiple points energy-dispersive x-ray spectroscopy (EDS) in a JEOL LSM-6500 scanning electron microscopy (SEM). Synchrotron powder x-ray diffraction (XRD) measurement was carried out in capillary transmission geometry using a Perkin Elmer amorphous silicon area detector at 28-ID-1 (PDF) beamline of the National Synchrotron Light Source II (NSLS II) at Brookhaven National Laboratory (BNL). The setup utilized a $\sim$74 keV ($\lambda$ = 0.16635 ${\AA}$) x-ray beam. Two dimensional diffraction data were integrated using Fit2D  software package \cite{Hammersley}. The Rietveld and PDF analysis were carried out using GSAS-II and PDFgui software packages, respectively \cite{Toby,Farrow}. Electrical resistivity, thermopower and thermal conductivity were measured in a quantum design PPMS-9 with standard four-probe technique. Sample dimensions were measured by an optical microscope Nikon SMZ-800 with resolution of 10 $\mu$m.

\section{RESULTS AND DISCUSSIONS}

\begin{table}[tbp]\centering
\caption{Atomic coordinates and displacements parameters for Ir$_{0.90(1)}$Sb$_{0.84(1)}$Se obtained from Rietveld refinement at 300 K in synchrotron powder XRD experiment.}
\begin{tabular}{ccccccc}
\hline\hline
Atom & occupancy & $x$ & $y$ & $z$ & $U_{\textrm{iso}}$({\AA}$^2$)\\
\hline
Ir & 0.904(4) & 0.0040(2) & 0.0040 & 0.0040 & 0.0034\\
Sb & 0.844(6) & 0.3790(3) & 0.3790 & 0.3790 & 0.0052(5)\\
Se & 1.000 & 0.6235(4) & 0.6235 & 0.6235 & 0.0010\\ \hline
\hline
\end{tabular}
\label{2}
\end{table}

\begin{figure}
\centerline{\includegraphics[scale=1]{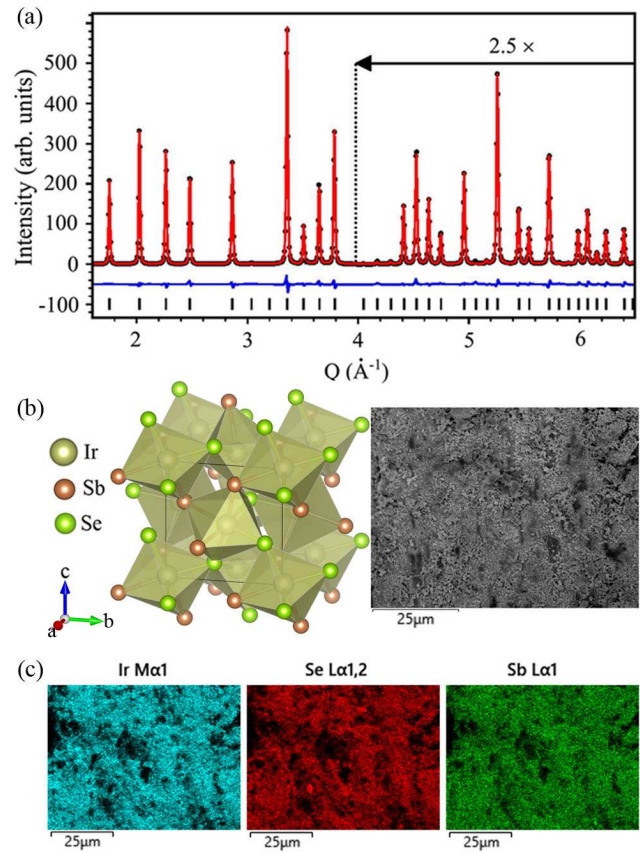}}
\caption{(Color online). (a) The Rietveld refinement of the background subtracted IrSbSe synchrotron powder x-ray diffraction. Plots show the observed (dots) and calculated (red solid line) powder patterns with a difference curve (blue). The black vertical tick marks represent Bragg reflections in the $P2_13$ space group. (b) Crystal structure and scanning electron microscopy (SEM) image along with (c) the energy dispersive spectroscopy (EDS) mapping on the IrSbSe sample.}
\label{XRD}
\end{figure}

Figure 1(a) shows the Rietveld refinement of synchrotron powder XRD for IrSbSe, indicating that all reflections can be well indexed in the $P2_13$ space group \cite{Hulliger}, and there are no extra impurity peaks. The determined lattice parameter $a$ = 6.20484(4) {\AA}. The crystal structure is a derivative of the structure of iron pyrite (FeS$_2$) \cite{Hulliger}, which can also be viewed as a network of distorted IrSb$_3$Se$_3$ octahedra with each Sb or Se belonging to three such octahedra [Fig. 1(b)]. The EDS mapping presented in Fig. 1(c) shows that the ratio of Ir : Sb : Se $\thickapprox$ 33(3) : 35(3) : 31(3), however, the Rietveld refinement indicates vacancy defects on both Ir and Sb atomic sites (Table I). The vacancy defect concentration is considerable, about 10\% on Ir and 15\% on Sb atomic sites, respectively.

Figure 2 shows the magnetic susceptibility of IrSbSe measured from 2 to 300 K in a magnetic field of $H$ = 10 kOe, which shows no evidence for long-range magnetic ordering below 300 K, and follows the Curie-Weiss behavior. The inverse susceptibility was modeled by the Curie-Weiss law, $\chi(T) = \chi_0 + C/(T-\theta_\textrm{W})$, where $C$ is the Curie constant, $\theta_\textrm{W}$ is the Weiss temperature, and $\chi_0$ = -8$\times10^{-5}$ emu mol$^{-1}$ Oe$^{-1}$ is the temperature-independent diamagnetic term. The linear fit in the inset of Fig. 2 gives $\theta_\textrm{W}$ = -3 K and $C$ = 1.67$\times10^{-3}$ emu K mol$^{-1}$ Oe$^{-1}$. The derived effective moment of $\mu_{\textrm{eff}}$ = $\sqrt{8C}$ $\approx$ 0.12 $\mu_\textrm{B}$/Ir is considerably smaller than the value of 2.83 $\mu_\textrm{B}$/Ir expected for spin-only $S$ = 1 with Ir$^{5+}$ (5$d^4$) ions. A similar magnetic behavior was also reported in Ir$^{5+}$ compounds NaIrO$_3$ and KIrO$_3$ \cite{NaIrO3,KIrO3}.

\begin{figure}
\centerline{\includegraphics[scale=1]{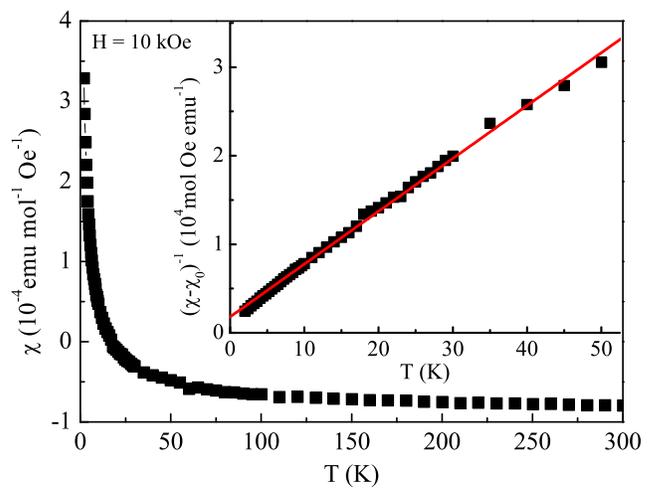}}
\caption{(Color online). Temperature dependence of magnetic susceptibility $\chi(T)$ for IrSbSe. The inset shows a Curie-Weiss plot with a fit to the Curie-Weiss law as explained in the main text.}
\label{XRD}
\end{figure}

\begin{figure}
\centerline{\includegraphics[scale=1]{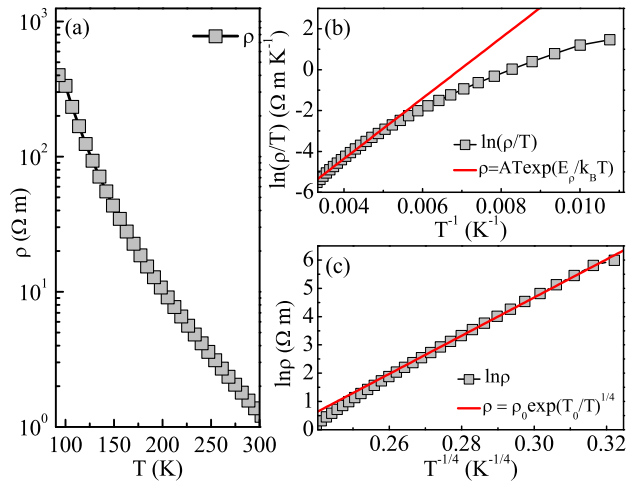}}
\caption{(Color online). (a) Temperature dependence of electrical resistivity $\rho(T)$ for IrSbSe. (b) ln($\rho/T$) vs $T^{-1}$ fitted by the adiabatic small polaron hopping model from 200 to 300 K. (c) ln$\rho$ vs $T^{-1/4}$ fitted by the Mott's variable-range hopping model from 100 to 200 K.}
\label{XRD}
\end{figure}

Temperature dependence of electrical resistivity $\rho(T)$ for IrSbSe is depicted in Fig. 3(a), showing an obvious semiconducting behavior. The value of $\rho_{\textrm{300K}}$ at room temperature is $\sim$ 1.2 $\Omega$ m. It is plausible to consider three typical models to describe the semiconducting behavior: (i) thermally activated model $\rho(T) = \rho_0 exp(E_\rho/k_\textrm{B}T)$, where $E_\rho$ is an activation energy; (ii) adiabatic small polaron hopping model $\rho(T) = AT exp(E_\rho/k_\textrm{B}T)$ \cite{Austin,YL1,YL2}; (iii) Mott's variable-range hopping (VRH) model $\rho(T) = \rho_0 exp(T_0/T)^{1/4}$, where $T_0$ is a characteristic temperature and is related to density of states available at the Fermi level $N(E_\textrm{F})$ and carrier localization length. To understand the transport mechanism in IrSbSe, it is necessary to fit the $\rho(T)$ data based on these three formulas. Figure 3(b) shows the fit result of the adiabatic small polaron hopping model. The extracted activation energy $E_\rho$ is $\sim$ 128(2) meV from 200 to 300 K, thus the band gap is estimated $\sim$ 256(4) meV. However, the $\rho(T)$ curve can also be well fitted using the thermally activated model (not shown here). With decreasing temperature, the $\rho(T)$ data from 100 to 200 K obeys the Mott's VRH model [Fig. 3(c)]. The derived characteristic temperature $T_0$ $\sim$ 2.1(1) $\times$ 10$^8$ K for the Mott's VRH conduction, and the corresponding localization length $\xi$ $\sim$ 1.1 {\AA} given by $\xi^3$ = $18/[k_\textrm{B}T_0N(E_\textrm{F})]$ \cite{Rong,Fried}, where $N(E_\textrm{F})$ is the density of states at the Fermi level.

\begin{figure}
\centerline{\includegraphics[scale=1]{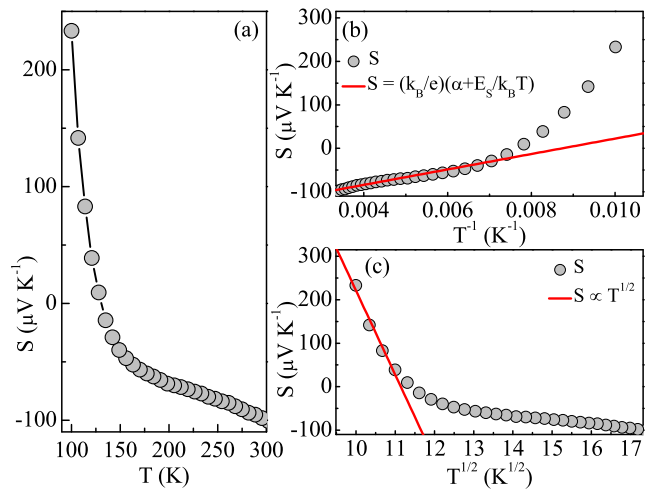}}
\caption{(Color online). (a) Temperature dependence of thermopower $S(T)$ for IrSbSe. (b) $S(T)$ vs $T^{-1}$ fitted by using $S(T) =
  (k_\textrm{B}/e)(\alpha+E_S/k_BT)$ from 200 to 300 K. (c) $S(T)$ vs $T^{1/2}$ with a linear fit from 100 to 121 K.}
\label{XRD}
\end{figure}

To distinguish the thermally activated model and polaron hopping model, we further measured temperature-dependent thermopower $S$ [Fig. 4(a)]. The $S(T)$ shows negative values in high temperature regime with a relative large value of $\sim$ -101 $\mu$V K$^{-1}$ at 300 K, indicating dominant electron-like carriers. With decreasing temperature, $S(T)$ changes its slope and the sign from negative to positive below 130 K, suggesting the hole-like carriers dominate at low temperatures. As shown in Fig. 4(b), the $S(T)$ vs $T^{-1}$ curve can be well fitted with the equation $S(T) = (k_\textrm{B}/e)(\alpha+E_S/k_\textrm{B}T)$ \cite{Austin}, where $E_S$ is an activation energy and $\alpha$ is a constant. The obtained activation energy for thermopower $E_S$ = 17.7(9) meV from 200 to 300 K [Fig. 3(b)], which is much smaller than that for conductivity $E_\rho$ = 128(2) meV [Fig. 3(b)]. This large discrepancy between $E_S$ and $E_\rho$ typically reflects the polaron transport mechanism of carriers. Within the polaron hopping model, $E_S$ is the energy required to activate hopping of carriers, while $E_\rho$ is the sum of energy needed for the creation of carriers and activating the hopping of carriers \cite{Austin}; therefore, $E_S$ is smaller than $E_\rho$. The weak temperature-dependent $S(T)$ at high temperatures also supports the small polaron conduction. With decreasing temperature, the resistivity evolves into the Mott's VRH dominant mechanism [Fig. 3(c)]. Within the Mott's VRH model the $S(T)$ can be described by $S(T) = S_0 + AT^{1/2}$ \cite{Zvyagin}; it can be seen from the rapid slope change of $S$(T), and $S$ varies as $T^{1/2}$ from 100 to 121 K [Fig. 4(c)].

\begin{figure}
\centerline{\includegraphics[scale=1]{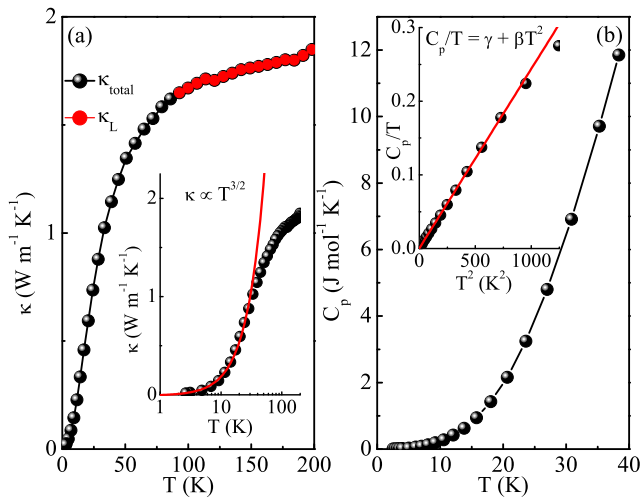}}
\caption{(Color online). Temperature dependence of (a) thermal conductivity $\kappa(T)$ and (b) heat capacity $C_p(T)$ for IrSbSe.}
\label{XRD}
\end{figure}

Figure 5(a) presents the temperature dependence of thermal conductivity $\kappa(T)$ for IrSbSe. Generally, $\kappa_{\textrm{total}} = \kappa_\textrm{e} + \kappa_{\textrm{L}}$, consists of the electronic charge carrier part $\kappa_\textrm{e}$ and the phonon term $\kappa_{\textrm{L}}$. The $\kappa_\textrm{e}$ estimated from the Wiedemann-Franz law is negligibly small due to large electrical resistivity of IrSbSe, indicating a predominantly phonon contribution. The $\kappa(T)$ shows a relatively low value of $\sim$ 1.85 W/K$\cdot$m at 200 K due to its structural complexity such as high nonstoichiometry. With decreasing temperature, $\kappa(T)$ follows a $T^{3/2}$-dependence below 30 K [inset in Fig. 5(a)]. This deviates from the common $\kappa$ $\sim$ $T^3$ usually observed in bulk crystals or thin films \cite{Toulokian,McConnell}, and it is close to power-law observed in thermal conductivity of nanostructures due to grain size variation \cite{WangZ,ZhaoH}. This might imply nanostructural differences that are induced by different vacancy defects in particular grains and associated phonon frequency changes. It should be noted that IrSbSe shows relatively low value of thermal conductivity $\sim$ 1.65 W/K$\cdot$m and thermopower of $\sim$ 0.24 mV/K around 100 K, yet poor electrical conductivity, leading to rather small values of power factor ($<$ 10$^{-2}$ $\mu$W/m$\cdot$K$^2$) and $zT$ ($<$ 10$^{-6}$). The specific heat $C_\textrm{p}(T)$ of IrSbSe at low temperatures is depicted in Fig. 5(b). By fitting the $C_\textrm{p}/T$ vs $T^2$ data below 30 K by using $C_\textrm{p}(T)/T = \gamma + \beta T^2$, we obtain the Sommerfeld electronic specific-heat coefficient $\gamma$ $\sim$ 0.4(2) mJ mol$^{-1}$ K$^{-2}$, as expected for an insulating ground state. The derived Debye temperature $\Theta_\textrm{D}$ = 288(1) K from $\beta$ = 0.244(1) mJ/mol$\cdot$K by using the equation $\Theta_\textrm{D} = [12\pi^4NR/(5\beta)]^\frac{1}{3}$, which implies an average sound velocity of $\nu_\textrm{s} \approx 2600$ m/s \cite{SV}. The phonon mean free path $l_\kappa \sim$ 0.5 $\mu$m at 2 K estimated from the heat capacity and thermal conductivity
via $\kappa_\textrm{L} = C_\textrm{p}\nu_\textrm{s}l_\kappa/3$ \cite{QH}.

\begin{figure}
\centerline{\includegraphics[scale=1]{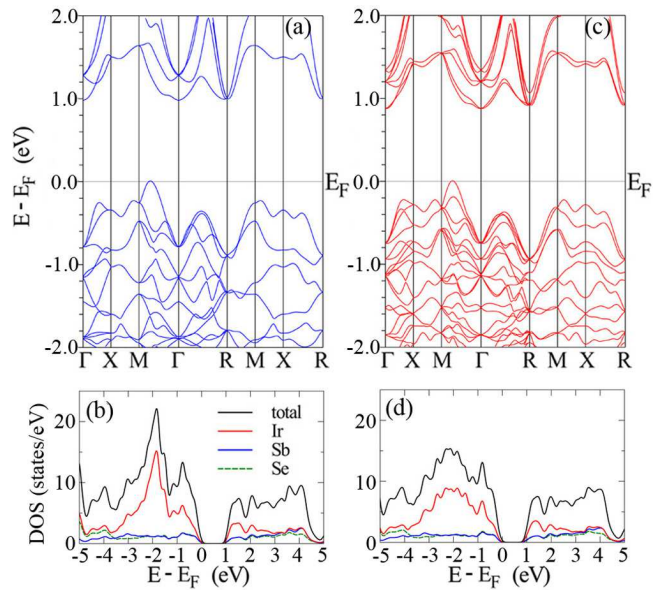}}
\caption{(Color online). (a) Band structure and (b) the atom-resolved density of states (DOS) calculated without spin-orbit coupling (SOC) on stoichiometric IrSbSe. (c) Band structure and (d) the DOS calculated with SOC.}
\label{XRD}
\end{figure}

We note that the polaronic transport is favored in materials with strong SOC that lack inversion symmetry \cite{GrimaldiC}. To get insight into the relevance of SOC, we further performed first-principles calculations using density function theory. We applied the WIEN2K \cite{43} implementation of the full potential linearized augmented plane wave method in the generalized gradient approximation using the PBEsol functional \cite{44} on stoichiometric IrSbSe \cite{Hulliger}. The SOC is treated in the second variation method. The basis size was determined by $R_\mathrm{mt}K_\mathrm{max}$ = 7 and the Brillouin zone was sampled with a regular $15\times 15 \times 15$ mesh containing 176 irreducible $k$ points to achieve energy convergence of 1 meV. As shown in Fig. 6(a-d), the calculated band structure and atom-resolved density of states (DOS) indicates that the system is a nonmagnetic semiconductor. There is a band gap of about 1 eV for the calculations without SOC [Fig. 6(a,b)] and about 0.9 eV for the case with SOC [Fig. 6(c,d)]. The total DOS (black line) rises rapidly from the band edges with the band character being mainly of Ir $5d$ orbitals. However, the contributions from Sb and Se $p$ orbitals are significant in reducing the Ir $5d$ orbital weight, compared with the dominant character of the transition metal in CoSbS and FeSbS \cite{46}. This weakens magnetic instability, if any, upon charge doping. The effects of SOC is more clearly seen in the band structures [Fig. 6(a,b)], where the SOC induces the band splitting is about 0.2 eV for the valence bands around the Fermi level. The band-structure plots also suggest that the valence and conduction bands are substantially massive. These are similar to the results for IrBiSe \cite{LiuZ}, where the SOC splitting of about 0.3 eV was reported. Thus, the split band in stoichiometric IrSbSe is also expected to be fully spin-polarized with 3D chiral spin texture \cite{LiuZ}. The present off-stoichiometric polycrystalline sample with a much smaller band gap suggests the occurrence of in-gap states.

\begin{figure}
\centerline{\includegraphics[scale=1]{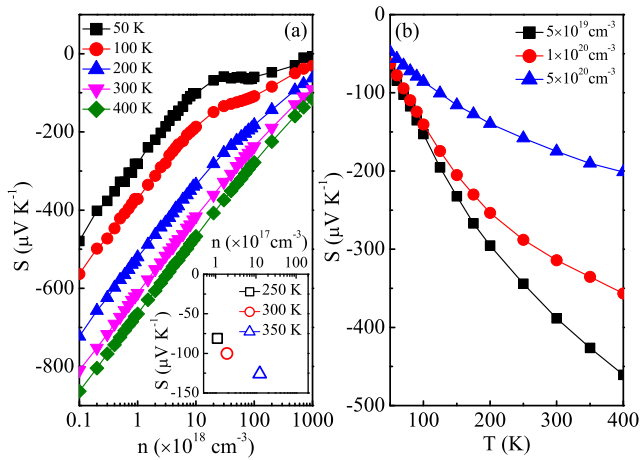}}
\caption{(Color online). Variation of calculated thermopower $S$ for electron doped case (a) as a function of carrier concentration $n$ for different temperatures and (b) as a function of temperature for different values of $n$. Inset in (a) shows the experimental $S$ and $n$ at the indicated temperatures.}
\label{XRD}
\end{figure}

In order to compare with the experimental thermopower ($S$), we calculate $S$ in the presence of SOC within the rigid band constant relaxation time approximation as implemented in the Boltztrap2 software package \cite{BT}. As shown in Fig. 7(a), the variation of $S$ with electron carrier concentration ($n$) for different temperatures shows Pisarenko behavior, i.e. logarithmic in carrier concentration, for a wide range of temperature and $n$. Such behavior has also been observed in many other semiconductors including CoSbS which lies in the same family \cite{46}. To make a more direct comparison with the experimental data, we also studied the temperature dependence of $S$ for different values of $n$ [Fig. 7(b)]. For this plot, we fix the value of $n$ at $T$ = 300 K and take into account the temperature dependence of $n(T)=\int_{E_c}^{\infty} dE g_c(E)/[e^{(E-\mu)/k_\textrm{B}T}+1]$ \cite{AshcroftMermin}, where $g_c$ is the density of states of conduction bands, $E_c$ is the conduction band edge, and $\mu$ is the chemical potential at that temperature. Temperature dependence of $\mu$ is obtained from the Sommerfeld relationship $\mu(T)=E_\textrm{F}-\pi^2k_\textrm{B}^2T^2g'(E_\textrm{F})/6g(E_\textrm{F})$. The calculated magnitude of
$S$ decreases as temperature decreases, in agreement with the experiment.

The Hall coefficient measured in the present polycrystalline sample is negative at the room temperature, in agreement with the negative value of $S$(300 K) [Fig. 4(a)]. The derived carrier concentration $n$(300 K) $\sim 1.9(1) \times 10^{17}$ cm$^{-3}$; it increases to $n$(350 K) $\sim 1.25(2) \times 10^{18}$ cm$^{-3}$ [inset in Fig. 7(a)]. For the values of $n$ calculated larger than the experimental ones, we find the magnitude as well as the temperature dependence of $S$ are similar to the experimental results. As shown in Fig. 8, our calculation can reproduce the linear dependence of $S$ with $T^{1/2}$ at low temperatures for the Mott's VRH dominant conduction, and $T^{-1}$ at high temperatures, respectively. The derived activation energy $E_S^{cal}$ = 65(5) meV, which is larger than the experimental value of $E_s$ = 17.7(9) meV but still much smaller than $E_\rho$ = 128(2) meV, confirming the polaronic nature at high temperatures. However, we should note that the experimentally observed $p-n$ transition at $T \sim$ 130 K has not been reproduced by the simulation, calling for further investigation of thermoelectric properties in this system beyond the constant relaxation time approximation by including temperature as well as energy and momentum dependent electron-phonon scattering rate. It is also plausible that such $p-n$ transition is the consequence of impurities in this system due to a large off-stoichiometry since it is not reported in a similar system CoSbS single crystal \cite{DuQ}. Further efforts on single crystal growth will be helpful to clarify this, and carrier optimization is needed to obtain better thermoelectricity performance.

\begin{figure}
\centerline{\includegraphics[scale=1]{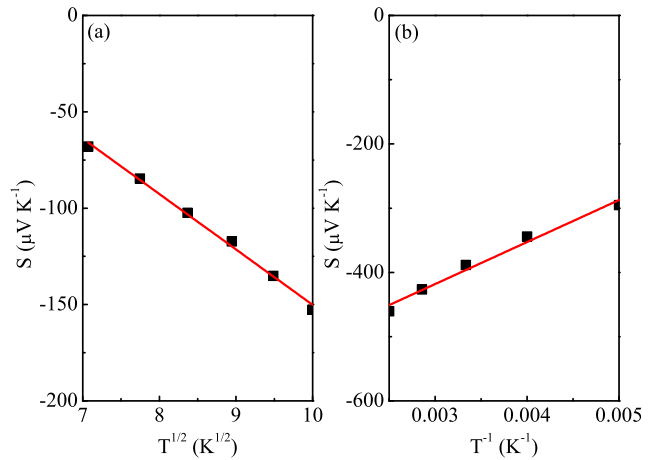}}
\caption{(Color online). Fitting of the calculated thermopower $S$ with $n$(300 K) = $5 \times 10^{19}$ cm$^{-3}$ in linear dependence of (a) $T^{1/2}$ and (b) $T^{-1}$ at low and high temperatures, respectively, similar to the experimental observation.}
\label{XRD}
\end{figure}

\section{CONCLUSIONS}

In summary, we observed poor electrical and thermal conductivity in IrSbSe polycrystal. Electronic transport mechanism indicates the polaronic nature at high temperatures and the Mott's VRH conduction at low temperatures. The thermopower is also simulated by using $ab-initio$ calculation. Structurally, IrSbSe crystallizes in the $P2_13$ space group even with high vacancy defect concentration, i.e. with Ir$_{0.90}$Sb$_{0.85}$Se stoichiometry. Such high defect tolerance implies possibilities for future carrier optimization by vacancy defects and chemical doping.

\section*{Acknowledgements}

Work at BNL is supported by the Office of Basic Energy Sciences, Materials Sciences and Engineering Division, U.S. Department of Energy (DOE) under Contract No. DE-SC0012704. This research used the 28-ID-1 beamline of the NSLS II, a U.S. DOE Office of Science User Facility operated for the DOE Office of Science by BNL under Contract No. DE-SC0012704. A portion of this work was performed at the National High Magnetic Field Laboratory, which was supported by the National Science Foundation Cooperative Agreement No. DMR-1644779 and the State of Florida.

$^{*}$Present address: Los Alamos National Laboratory, Los Alamos, NM 87545, USA.\\

\end{document}